\documentclass[12pt, draftclsnofoot, onecolumn]{IEEEtran}
\usepackage{amsfonts}
\usepackage{amssymb}
\usepackage{amsthm}
\usepackage{amsmath,amsfonts,amssymb}
\usepackage[dvips]{graphicx}
\usepackage{verbatim}
\usepackage{bm}
\usepackage[ruled,vlined]{algorithm2e}
\usepackage{cite}
\usepackage{color}

\IEEEoverridecommandlockouts

\begin{document}

\title{Full-Duplex Millimeter-Wave Communication}

\author{Zhenyu Xiao,~
        Pengfei Xia,~
        Xiang-Gen Xia
%and Xiang-Gen Xia
\thanks{Zhenyu Xiao is with Beihang University.}
\thanks{Pengfei Xia is with the Key Laboratory of Embedded System and Service Computing, Tongji University.}
\thanks{Xiang-Gen Xia is with University of Delaware.}
}

\maketitle
\begin{abstract}
The potential of doubling the spectrum efficiency of full-duplex (FD) transmission motivates us to investigate FD millimeter-wave (FD-mmWave) communication. To realize FD transmission in the mmWave band, we first introduce possible antenna configurations for FD-mmWave transmission. It is shown that, different from the cases in micro-wave band FD communications, the configuration with separate Tx/Rx antenna arrays appears more flexible in self-interference (SI) suppression while it may increase some cost and area versus that with the same array. We then model the mmWave SI channel with separate Tx/Rx arrays, where a near-field propagation model is adopted for the line-of-sight (LOS) path, and it is found that the established LOS-SI channel with separate Tx/Rx arrays also shows spatial sparsity. Based on the SI channel, we further explore approaches to mitigate SI by signal processing, and we focus on a new cancellation approach in FD-mmWave communication, i.e., beamforming cancellation. Centered on the constant-amplitude (CA) constraint of the beamforming vectors, we propose several candidate solutions. Lastly, we consider an FD-mmWave multi-user scenario, and show that even if there are no FD users in an FD-mmWave cellular system, the FD benefit can still be exploited in the FD base station. Candidate solutions are also discussed to mitigate both SI and multi-user interference (MUI) simultaneously.
\end{abstract}


%Antenna configuration: shared VS separate, circulator, SI strength and flexibility, etc.

% SI channel the packaging, reflection effect.

% analog beamforming

%\IEEEpeerreviewmaketitle
\section{Introduction}
\IEEEPARstart{M}{illimeter-wave} (mmWave) communication is a promising technology for next-generation wireless communication \cite{xia2016robust,alkhateeb2014mimo,sun2014mimo,xiao2016codebook} owing to its abundant frequency spectrum resource, which promises a much higher capacity than the existing micro-wave band wireless local area networks (WLANs) and cellular mobile communications.
In order to bridge the link budget gap due to the extremely high path loss in the mmWave band, beamforming with large antenna arrays are generally required. Subject to expensive radio-frequency (RF) chains, mmWave communication adopts particular beamforming structures, i.e., analog beamforming/combining is usually preferred for one-stream transmission, where all the antennas share a single RF chain and have constant-amplitude (CA) constraint on their weights \cite{xia2016robust,xiao2016codebook}. Meanwhile, hybrid analog/digital precoding/combining has also been proposed to realize multi-stream/multi-user transmission \cite{alkhateeb2014mimo,sun2014mimo}, where a small number of RF chains are used together with a large antenna array with much more number of antenna elements.

On the other hand, when applying mmWave communication in practice, its benefit may be offset by the conventional time division duplex (TDD) or frequency division duplex (FDD). In particular, with TDD sufficient Tx/Rx switching guard time should be reserved to make sure the Tx/Rx circuits work normally. Moreover, when considering multiple access, there may be significant protocol overhead \cite{Xia_2011_60GHz_Tech}. Even for bi-directional point-to-point transmission, there is still a lot of handshaking overhead  \cite{Xia_2011_60GHz_Tech}. While with FDD a large guard band should be arranged to make sure the interference leakage is small enough. In brief, the conventional TDD/FDD may reduce the effective capacity of mmWave communications. For this reason, we explore full duplex (FD) for mmWave communications in this paper. FD has received significant interest in the past decade. Different from TDD/FDD, transmission and reception occur in the same time/frequency resource block in FD communication, which may double the spectrum efficiency \cite{bharadia_2013,Sabharwal2014JSAC}. Moreover, FD provides more efficient and flexible access strategies for multiple access \cite{Sabharwal2014JSAC}, and may not need handshaking for bi-directional point-to-point transmission, which can further increase the practical efficiency of mmWave communication.

However, FD-mmWave communication faces particular challenges. Naturally, the most critical issue in FD-mmWave is also self-interference (SI), which is the transmitted signal received by the local receiver at the same node and needs to be cancelled. The SI itself as well as SI cancellation in FD-mmWave communication has its own features. Firstly, unlike the micro-wave band communications, large antenna arrays are generally required in mmWave communication, which means that specific antenna settings are required to enable FD-mmWave transmission. This is studied in Section II, {where we show that the configuration with separate Tx/Rx arrays appears more flexible in SI suppression, while it may increase some cost and area.} As using antenna Tx/Rx arrays leads to a matrix SI channel, we then model the line-of-sight (LOS) component of the SI channel by exploiting a near field signal propagation model in Section III, and we show that the LOS-SI channel also demonstrates spatial sparsity when the number of antenna elements is large. Based on the SI channel, we next explore signal processing schemes to mitigate SI in Section IV, where the spatial sparsity of the SI channel is exploited. In particular, we propose a new SI cancellation method termed as beamforming cancellation, which utilizes the beamforming functions of the Tx and Rx antenna arrays. In Section V, we further consider FD-mmWave communication in a multi-user scenario, and show that the FD benefit can be exploited in the FD base station even if there is no FD user. We also discuss possible solutions to mitigate both SI and multi-user interference (MUI) at the same time. Lastly, we conclude the paper in Section VI.

\section{Antenna Configuration}

\begin{figure}[t]
\begin{center}
  \includegraphics[width=16 cm]{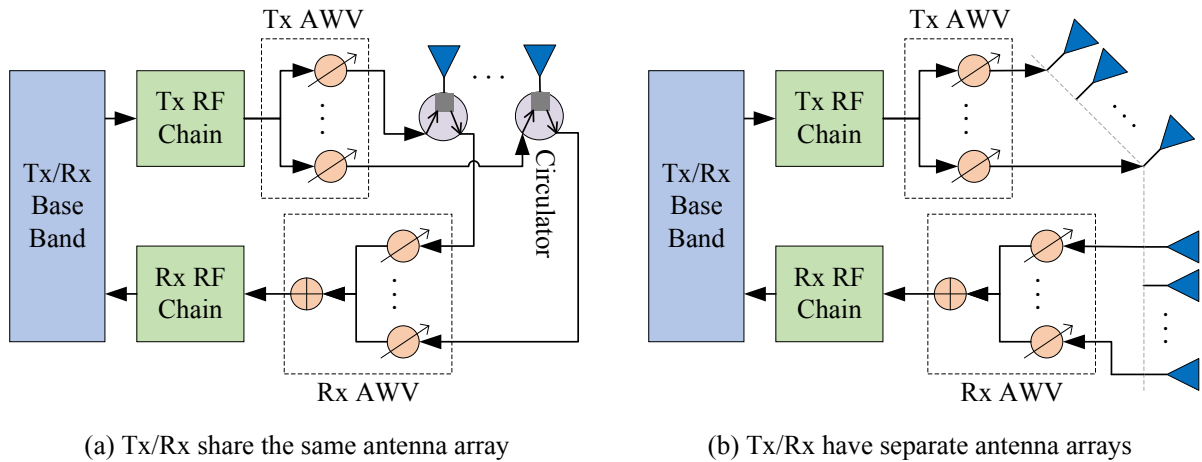}
  \caption{Possible antenna configurations of an FD-mmWave node.}
  \label{fig:analog}
\end{center}
\end{figure}

The first issue that we need to address is antenna configuration for FD-mmWave communication. In the mmWave band, the high power consumption of mixed signal components and expensive radio-frequency (RF) chains make it difficult to realize digital baseband beamforming as used in the conventional multiple-input multiple-output (MIMO) systems. Consequently, an analog beamforming/combining structure \cite{xia2016robust,xiao2016codebook} is generally adopted\footnote{A hybrid analog/digital precoding/combining structure \cite{alkhateeb2014mimo,sun2014mimo} is also frequently used for multi-stream/multi-user transmission.}. Hence, to enable FD transmission in the mmWave band, we introduce two possible antenna configurations as shown in Fig. \ref{fig:analog}. There are one Tx RF chain and one Rx RF chain in both configurations. In Fig. \ref{fig:analog}--(a) Tx/Rx share the same antenna array by using circulators. The function of the circulator is to separate the Tx signal path from the Rx signal path, and meanwhile let the Tx/Rx signals be transmitted/received through the antenna connecting to it. In Fig. \ref{fig:analog}--(b) Tx/Rx are equipped with separate antenna arrays. We compare these two configurations as follows.

{Firstly, we compare the capability of SI suppression between the two configurations. For the configuration of sharing the same antenna array in Fig. \ref{fig:analog}--(a), SI is suppressed by a circulator. The isolation of a circulator is dependent on the quality of the circulator. In the mmWave band, the isolation is still limited in general and needs further improvement. For instance, the isolation of the circulator designed in \cite{kijsanayotin2014millimeter} is only 18 dB, less than that in the micro-wave band in general. On the other hand, even if the circulator is good enough, the signal transmitted from an arbitrary antenna element will also result in significant SI to its neighbor antenna elements, because the distances between these antenna elements are quite small, basically at a level of the wavelength of the carrier. In contrast, with separate antennas as that in Fig. \ref{fig:analog}--(b), SI can be suppressed by blocking the signals between Tx/Rx antenna arrays and/or locating the Tx/Rx antennas far away from each other. Moreover, with separate arrays, the parameters of Tx/Rx array positioning may also be exploited as additional degrees of freedom to mitigate SI and optimize the system performance, as will be discussed in detail later.}

{Then, we compare the cost and area of the two configurations. It is clear that the configuration with separate Tx/Rx arrays requires an additional array, while it does not need circulators. It is hard to say whether an antenna array or the circulators have higher cost. Even if an antenna array has higher cost, it is affordable in general because it is only a small part of the total cost of the whole transceiver. In addition to cost, area is also important for an FD device, especially for portable devices. Note that in the micro-wave band the size of a micro-wave antenna is relatively large \cite{bharadia_2013}. Hence, using sperate Tx/Rx antennas will double the antenna area, which may be unfavored for small devices, while sharing the same antennas can save a large area. In contrast, in the mmWave band the size of a mmWave antenna is small; thus an additional antenna array may occupy only a small area. For instance, for a $4\times 4$ uniform planar array (UPA) with half wavelength spacing at 30 GHz, the required area is only about $2\times 2~\rm{cm}^2$. However, when the number of the antenna elements is too large, the required area may be also large.}

%Hence, although the configuration with separate Tx/Rx arrays has higher cost and a larger size, it may be affordable for an FD-mmWave device in general.

{In summary, the configuration with separate Tx/Rx arrays appears more flexible in SI suppression over the one sharing the same array for FD-mmWave communication, while it may increase some cost and area for an FD-mmWave device. In the rest of this paper, we will use the configuration with separate Tx/Rx arrays for FD-mmWave communication, while the one with the same array can be seen as a particular case of the separate arrays from the signaling viewpoint, i.e., the two arrays completely coincide with each other.}

%For these reasons, we prefer the configuration with separate arrays for FD-mmWave communication, and will discuss channel modeling and SI cancellation issues with this configuration in the following.

\section{SI Channel Modeling}
%In this section, we will first show how much stronger SI could be than the background noise in mmWave band, and then model the mmWave SI channel. To do so, we compute a typical value of SI observed at the Rx array in an FD-mmWave node with separate Tx/Rx arrays, where a uniform linear array (ULA) with half wavelength antenna space is adopted. If the wavelength of the carrier frequency is 10 mm, the signal bandwidth is 100 MHz, and the transmission power (per antenna) is 5 dBm, according to the Friis formula, the SI (without antenna cancellation) observed at an antenna 100 mm ($10\lambda$) away is $5-20\log_{10}(4\pi\times 10)=-37$ dBm, which is much greater than a typical noise power $\sigma^2=10\log_{10}(\kappa TB)=10\log_{10}(1.38\times 10^{-23}\times 300\times 10^8\times 10^3)=-93.83$ dBm, where $\kappa,~T,~B$ are the Boltzmann constant, ambient temperature and bandwidth, respectively. When Tx/Rx shares the same array, the SI will be even much stronger. Hence, SI should be taken into account in FD-mmWave communication, and to mitigate the SI we need to know about the SI channel.

An mmWave SI channel refers to the channel from the local Tx array to the local Rx array at the same node. It is different from {a mmWave} communication channel between two different nodes. For comparison, we briefly introduce {a mmWave} communication channel, which is expected to have limited scattering, and multi-path components (MPCs) are mainly generated by a few reflections, i.e., {a mmWave} communication channel has spatial sparsity. Different MPCs have different angles of departure (AoDs) and angles of arrival (AoAs) \cite{alkhateeb2014mimo,sun2014mimo}. In particular, the communication channel from Node 1 to Node 2 can be expressed as $
{\bf{H}}_{\rm{CM}} = \sum_{\ell  = 1}^L {{\lambda _\ell }{\bf{a}}({N_{\rm{AN2}}},{\Omega _\ell }){\bf{a}}{{({N_{\rm{AN1}}},{\psi _\ell })}^{\rm{H}}}}$,
where $\lambda_\ell$ is the complex coefficient of the $\ell$-th path, $L$ is the number of MPCs, ${\bf{a}}(\cdot)$ is the \emph{steering vector function}, ${N_{\rm{AN1}}}$ and ${N_{\rm{AN2}}}$ are the numbers of antennas at Node 1 and Node 2, respectively, ${\Omega _\ell }$ and ${\psi _\ell }$ are cos(AoD) and cos(AoA) (called AoA or AoD in the remaining of this paper) of the $\ell$-th path, respectively. ${\bf{a}}(\cdot)$ depends on the antenna array at Tx/Rx, and if uniform linear arrays (ULAs) with half wavelength spacing are adopted, it can be defined as
${\bf{a}}(N,\Omega ) =\frac{1}{{\sqrt N }}[e^ {{\rm{j}}\pi 0\Omega},~e^{ {\rm{j}}\pi 1\Omega },...,e^{{\rm{j}}\pi (N - 1)\Omega}]^{\rm{T}}$,
where $N$ is the number of antennas at the Tx or Rx, $\Omega$ is AoD or AoA. This model implicitly assumes that the distance of the two nodes is far greater than the wavelength of the carrier, such that a far-field model is adopted.

In contrast, the far-field range condition, i.e., $R\geq 2D^2/\lambda$ \cite{fenn1990evaluation}, does not hold in general for the LOS component of mmWave SI, where $R$ is the distance between Tx and Rx arrays, $D$ is the diameter
of the antenna aperture, $\lambda$ is the wavelength
of the carrier.
For instance, considering a half-wavelength spaced ULA with 32 elements,
the far-field range should satisfy
$R\geq 2(16\lambda)^2/\lambda=1024\lambda$, which is basically too large for small-size devices like mobile phones
or laptops even at the mmWave band. Thus, the SI channel may
have to use the near-field model, which has a spherical wavefront
\cite{fenn1990evaluation}.
In such a case, the LOS path of the SI channel is highly dependent on the structures and relative positions of the Tx/Rx antenna arrays.

\begin{figure}[t]
\begin{center}
  \includegraphics[width=10 cm]{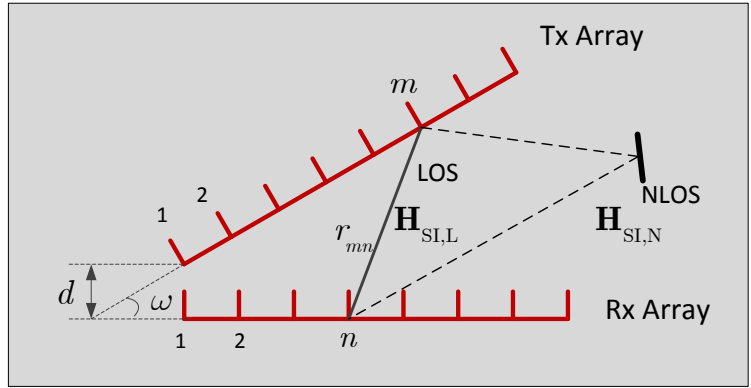}
  \caption{An illustration of Tx/Rx array positioning.}
  \label{fig:channel}
\end{center}
\end{figure}

In addition to the LOS component, there are also non-LOS (NLOS) components for the mmWave SI channel, due to the possible reflectors near the mmWave device ( see Fig. \ref{fig:channel}), i.e., ${\bf{H}}_{\rm{SI}}={\bf{H}}_{\rm{SI,L}}+{\bf{H}}_{\rm{SI,N}}$. Different from the LOS component, where the direct path between Tx/Rx antennas is short, the propagation distances via the NLOS paths are much longer, and can satisfy the far-field range condition in general \cite{Li2014feasibility}. Hence, the NLOS components may adopt the same model as the mmWave communication channel as above introduced. Moreover, compared with the LOS components, the NLOS components experience much higher propagation loss as well as additional reflection loss. Hence, the strength of the NLOS components is much weaker than that of the LOS component. For these reasons, we put more attention to the LOS component of the mmWave SI channel.

Without loss of generality, we consider an antenna placement
as shown in Fig. \ref{fig:channel}.
The distance between the first elements of the two arrays is $d$, and the angle between these two ULAs is $\omega$.
With this antenna placement, {the channel gain} between the $n$-th receive antenna and the $m$-th Tx antenna, i.e., the coefficient corresponding to the $n$-th row and the $m$-th column of ${\bf{H}}_{\rm{SI,L}}$ (the SI channel matrix with a near-field model) is \cite{liaodesign2015,lin2006near} $
[{\bf{H}}_{\rm{SI,L}}]_{nm} =
{h_{nm}} = \frac{\rho }{{{r_{nm}}}}\exp \left( { - j2\pi \frac{{{r_{nm}}}}{\lambda }} \right)$,
where ${r_{nm}}$ is the distance between the $m$-th element of the transmit array and the $n$-th element of the receive array, and $\rho$ is a constant for power normalization.

%On the other hand the SI channel has an asymptotical low-rank feature. In other words, as the distance of the Tx/Rx arrays increases, the SI channel gradually appears the low-rank feature. To illustrate this, let us see, as shown in the left figure in Fig. \ref{fig:channel_order}, the summation of the two maximal eigenvalues of ${\bf{H}}_{\rm{SI}}{\bf{H}}_{\rm{SI}}^{\rm{H}}$ given that ${\bf{H}}_{\rm{SI}}$ is normalized with $\rm{tr}({\bf{H}}_{\rm{SI}}{\bf{H}}_{\rm{SI}}^{\rm{H}})=1$. We can find that most channel power is distributed in only a few principal-eigenvector spaces. As $d/\lambda$ increases, the distribution will be more concentrated on the most principal-eigenvector space. When $d/\lambda$ approach infinity, the SI channel will become a rank-1 channel, and the near-field model will become a far-field model. When $d/\lambda$ is small, however, whether the distribution of the channel power is concentrated or not depends on $\omega$. More importantly, when $d/\lambda$ is not very large, the SI channel does not have spatial sparsity, which is different from the communication channel. To show this, let us see the right figure of Fig. \ref{fig:channel_order}, where the channel gains are obtained by setting Rx AWV as steering vectors with (cosine) steering angles from -1 to 1 and Tx AWV a random normalized vector.

\begin{figure}[t]
\begin{center}
  \includegraphics[width=18 cm]{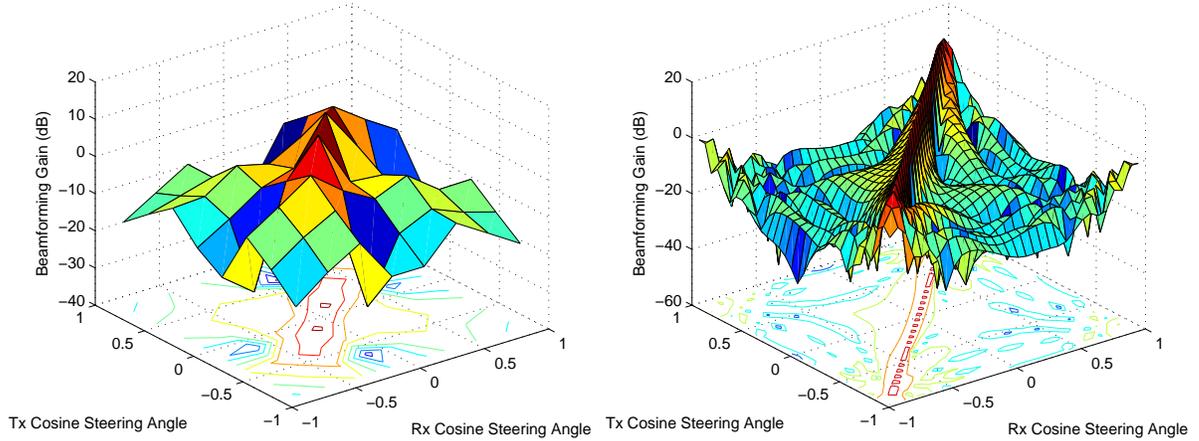}
  \caption{Spatial feature of the mmWave LOS-SI channel, where $\omega=0$. The number of antenna elements is 8 for both the Tx/Rx arrays in the left hand side figure, while 32 for the right hand side figure.}
  \label{fig:MAG_Channel}
\end{center}
\end{figure}

Interestingly, although the LOS-SI channel exploits the near-field model, from our model we find that when the number of antenna elements is large, the LOS-SI channel also shows spatial sparsity, which is similar to the mmWave communication channel. To show this feature, we set the Tx/Rx {antenna weight vectors (AWVs)} as steering vectors with (cosine) steering angles from -1 to 1, {and obtain the beamforming gains $|{\bf{a}}(N,\alpha)^{\rm{H}}{\bf{H}}_{\rm{SI,L}}{\bf{a}}(N,\beta)|$ with these angle combinations. Fig. \ref{fig:MAG_Channel} shows the results, where we can find that when the number of antenna elements is larger, the angle area, in which the beamforming gain is significant w.r.t. the largest beamforming gain, is smaller. When the number of elements is 32, as shown in the right hand side figure, the LOS-SI channel shows distinct spatial sparsity. In particular, the beamforming gain is significant only when $\beta-\alpha\thickapprox \omega$, the angle of the Tx/Rx arrays; otherwise it is small.} This result implies that most of the SI power is distributed along only a few Tx/Rx steering angle pairs. Hence, by steering the Tx/Rx AWVs to other angle pairs, the SI can be efficiently mitigated provided the number of antennas is sufficiently large, and the parameters, e.g., $\omega$, can also be optimized for SI cancellation as will be shown later.

As we can see, {a mmWave} SI channel can be very complicated, which includes both LOS component with a near-field propagation model and NLOS components with a far-field propagation model. Fortunately, due to the packaging of a portable FD-mmWave device, the propagation circumstances within the device are basically stable. For instance, a mobile phone can be packaged with a metal shell, which can block the reflected mmWave signals from the back of the mobile phone and thus prevents the outside objects, e.g., hands, from affecting the SI channel. Hence, in general {a mmWave} SI channel can be seen slow-varying. As a consequence, we usually have enough time to make an accurate mmWave SI channel estimation.

%%\subsection{SI Channel Estimation}
%In an FD-mmWave communication system, both the SI channel and communication channel need to be estimated for beamforming and combining. As a mmWave communication channel has the feature of directivity and is sparse in the angle domain, schemes like beam searching \cite{hur2013millimeter}, iterative training \cite{Xiaozy2014BeamTrain}, and compressed sensing \cite{alkhateeb2014channel} can be adopted for the communication channel estimation. However, these methods are not viable for mmWave SI channel estimation, because the mmWave SI channel do not have the feature of directivity and spatial sparsity.
%
%In fact, mmWave SI channel estimation can be more straightforward that mmWave communication channel estimation. For instance, one can estimate a scalar channel coefficient between a single Tx/Rx antenna pair (i.e., the $i$-th antenna and the $j$-th antenna) once a time. After $N_{\rm{AN}}^2$ measurements, the SI channel matrix can be estimated. Although $N_{\rm{AN}}^2$ is large in mmWave communication since $N_{\rm{AN}}$ is large, each measurement may require only one symbol thanks to the high strength of SI, rather than a long training sequence like those in \cite{hur2013millimeter,Xiaozy2014BeamTrain,alkhateeb2014channel} for communication channel estimation. Considering that mmWave channel is in general slow-varying, the time cost of the SI channel estimation is yet affordable, and we may have enough time to make an accurate SI channel estimation.

\section{SI Cancellation}

\subsection{SI Cancellation in FD-mmWave Communication}
As SI is still significant in FD-mmWave communication even with separate Tx/Rx antenna arrays, we need to mitigate it as much as possible. In the existing micro-wave band FD systems, antenna cancellation, which suppresses signals between Tx/Rx antennas, RF cancellation, which subtracts SI in the RF domain, and baseband cancellation, which mitigates residual SI in the baseband, are three typical methods to cancel SI, and a combination of them usually achieves better performance \cite{Sabharwal2014JSAC,bharadia_2013}.

For FD-mmWave communication, SI cancellation has its particular challenges. Due to the high frequency and wide band signals, the RF impairments are usually more significant than those in the micro-wave band, including power amplifier (PA) nonlinearity, in phase-quadrature (IQ) mismatch, phase noise, etc. These RF impairments will be embedded in SI and can only be effectively mitigated in RF and antenna cancellations \cite{Sabharwal2014JSAC}, since to cancel these impairments in the baseband is difficult due to the unknown impairment parameters. Unfortunately, in mmWave communications, RF design faces particular challenges due to the high frequency and wide band signals, which means that the parameter design for RF cancellation may be inaccurate, and thus probably only a small part of SI can be reduced by RF cancellation. Moreover, within a small portable devices Tx/Rx antenna cancellation can also only mitigate limited SI. Hence, FD-mmWave communication must seek for other efficient SI cancellation methods.

%antenna cancellation is particularly favored, because mmWave signals are relatively easier to be blocked, and baseband cancellation is also applicable. In contrast, RF cancellation may be not as suitable as the others, because for one reason it requires an additional RF chain which is both power and cost expensive, and for another reason the performance of RF cancellation may be not good since there may be more non-ideal issues in mmWave RF design.

Fortunately, in addition to the three typical SI cancellation methods, the inherent beamforming structure of mmWave communication enables another important approach to mitigate SI, i.e., beamforming cancellation. In fact, without considering the CA constraint of mmWave communication, the beamforming technology can even completely cancel SI, including Tx RF impairments, by using the zero-forcing (ZF) filtering. In \cite{Liu2016FDmmWave} it is shown that with unconstrained beamforming, FD-mmWave communication can achieve promising performance. However, under the CA constraint, the SI may not be completely mitigated, and it becomes a challenge for FD-mmWave communication to mitigate SI under the CA constraint. We focus on this issue in the following subsections.

%In regular mmWave communication, beamforming/combining are used to achieve array gain to compensate for the high propagation loss in the mmWave band \cite{alkhateeb2014mimo,roh2014millimeter,sun2014mimo}. In FD-mmWave communication, beamforming/combining has an additional task in addition to achieving array gain, i.e., SI mitigation. Hence, in FD-mmWave communication we face the challenge of SI driven beamforming and combining.

\subsection{Constant-Amplitude Beamforming Cancellation}
Subject to the CA constraint, in regular mmWave communication analog beamforming/combining is generally realized by steering the Tx AWV and Rx AWV to the AoD and AoA of the strongest MPC, such that high array gain can be achieved. However, this method may not be viable in FD-mmWave communication, where strong SI may exist. In other words, the conventional Tx/Rx AWV settings may not be able to effectively reduce SI, although they can achieve a higher array gain. As a result, the overall signal-to-interference plus noise power ratio (SINR) may be not high, as pointed out in \cite{Li2014feasibility}. Hence, in FD-mmWave communication SI driven beamforming/combining must be considered. The problem of CA beamforming cancellation (CA-BFC) is challenging, because the objective function, e.g., the {achievable sum rate (ASR)} of the two nodes in an FD link, is usually complicated and non-concave due to the SI. Moreover, the CA constraints of the Tx/Rx AWVs are also non-concave/convex.
%Compared with the conventional Tx/Rx joint beamforming problem in MIMO scenarios, there are CA constraints in CA-BFC; while compared with the regular mmWave beamforming problem, there are strong SI in CA-BFC.

%From the formulation we can also see why the low-rank feature of the SI channel is favored. Intuitively, the lower the rank of the SI channel is, the smaller number of vector spaces are ``contaminated'' by the SI. Thus, the Tx/Rx AWVs have a larger number of degrees of freedom to maximize the sum achievable rate. When Tx/Rx arrays are separated far enough, the SI channel will be a rank-1 channel, and it will also have the features of directivity and the spatial sparsity. As a consequence, even the conventional beam steering method may achieve good performance. However, under the typical case that the distance of Tx/Rx arrays is small, we must consider how to reduce SI by more effective beamforming methods.

While it is difficult to find an optimal solution of CA-BFC, it is interesting and with significance to find suboptimal solutions with low computational complexity. Here we discuss some preliminary solutions. Intuitively, as there are only phase uncertainty for each element of the Tx/Rx AWVs under the CA constraints, we may search from 0 to $2\pi$ for each element. Theoretically, provided small enough step size, an optimal solution can be found. However, as the numbers of antennas are large in general, the exhaustive search method is basically infeasible due to exhibitively high computational complexity. Hence, we propose the following two candidate methods:
\begin{itemize}
  \item \textbf{Method 1:} To restrict the Tx/Rx AWVs to the steering vector space, i.e., we assume that the Tx/Rx AWVs are all steering vectors and have uncertainty only on the steering angles. Under this assumption, we can search over the steering angle spaces of the Tx/Rx AWVs and find the best steering angles for them to optimize the object. This method is viable just because that the mmWave SI channel also has spatial sparsity, as shown in Section III.
  \item \textbf{Method 2:} To first find AWVs by solving the CA-BFC problem without considering the CA constraints, and then find AWVs that satisfy the CA constraints and have the minimal Euclidean distances from the already found AWVs.
\end{itemize}
%
%There may be also other methods to deal with the CA constraint problem. For instance, we can use semi-definite relaxation to convert the CA constraint to a linear constraint, and try to relax the CA-BFC problem to a convex problem. Although this method may be also feasible, the computational complexity is too high for real-time online computation.

\begin{figure}[t]
\begin{center}
  \includegraphics[width=16 cm]{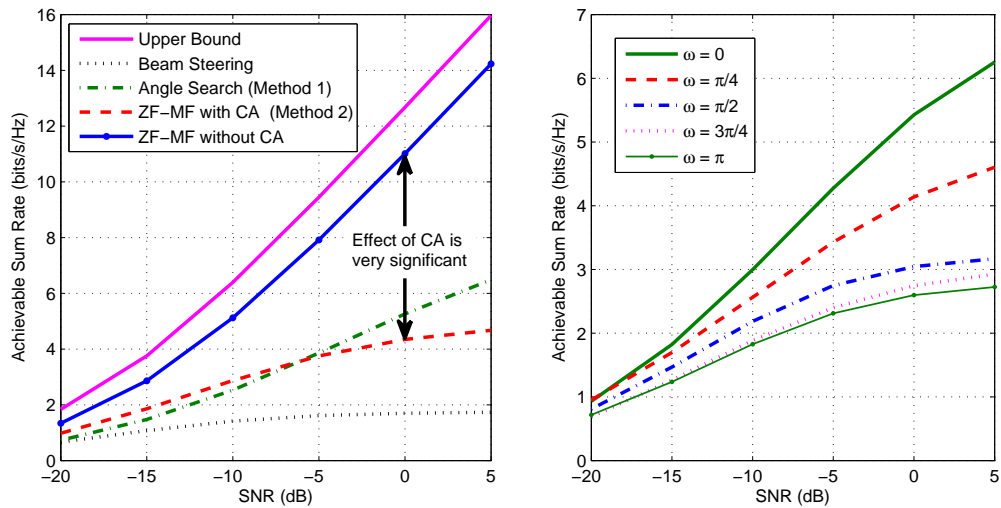}
  \caption{\textbf{Left:} Achievable sum rate comparison between different beamforming schemes in an FD-mmWave communication system, {where for each scheme the numbers of Tx and Rx antennas are 16 and 8, respectively, $\omega=\pi$ and $d/\lambda=5$}. \textbf{Right:} Achievable sum rate comparison of Angle Search with varying $\omega$, where the number of Tx and Rx antennas is 16 at both nodes, $d/\lambda=1$. In both figures the signal/noise powers are the same at the two nodes, and the SI is 25 dB w.r.t. the signal power.}
  \label{fig:analog_cmp}
\end{center}
\end{figure}

%Comparison figure: AR, a bound, direct beamforming, direct search, the gap can be achieved via different approaches.
The left hand side figure of Fig. \ref{fig:analog_cmp} shows ASR comparison between different schemes in an FD-mmWave communication system. The upper bound is obtained by assuming that SI is zero and there are no CA constraints. The ``Beam Steering'' method is to simply set Tx/Rx AWVs to steer along {a strong MPC} without considering the SI, which is a typical beamforming method used for regular mmWave communication. ``Angle Search'' refers to Method 1. To use Method 2, we adopt ZF-Matched Filter (ZF-MF) to obtain the beamforming vectors, and then add the CA constraints according to Method 2. ZF-MF is a linear beamforming algorithm which performs MF at the Tx first and then performs ZF at the Rx.
%With ZF-MF, the SI can be completely cancelled. However, the operation of adding the CA constrains according to Method 2 may lead to SI.

From the left hand side figure of Fig. \ref{fig:analog_cmp} we can observe that a conventional beamforming approach (Beam Steering) which does not take SI into account cannot achieve a satisfactory performance. In addition, although the simple Angle Search and ZF-MF with CA methods are viable, there is a significant performance gap from the upper bound, which means that these two methods may better work in low SI cases; when SI is strong (as shown in the figure), these two methods need to work together with other SI cancellation methods, like antenna cancellation or baseband cancellation. Meanwhile, more efficient methods are in demand to deal with the SI under the CA constraint. In fact, it is just the CA constraint that limits the performance of Method 2. Without the CA constraint, ZF-MF can achieve a performance close to the upper bound.

On the other hand, we've mentioned in Section III that the SI channel also has spatial sparsity, which can be exploited to design methods to mitigate SI, e.g., the Angle Search method. As we can expect, the Tx/Rx array positions may significantly affect the performance of this category of methods. The right hand side figure of Fig. \ref{fig:analog_cmp} shows the ASR comparison of Angle Search with varying $\omega$. It is clear that $\omega$ can dramatically affect the performance, and thus may be considered as another degree of freedom to optimize the ASR. It is noteworthy that different methods may favor different $\omega$.

\subsection{Beamforming Cancellation with Double RF Chains}
% Fig. 4
We've seen that the CA constraints greatly limit the performance of beamforming cancellation. If we can find an approach to remove or avoid the CA constraints, beamforming cancellation can become very promising even without the need to work together with other SI cancellation methods \cite{Liu2016FDmmWave}. Fortunately, the work in \cite{sohrabi2015hybrid} provides a great opportunity to realize this. It is shown in \cite{sohrabi2015hybrid} that an arbitrary vector ${\bf{v}}$ can be expressed as the linear combination of two CA vectors, i.e., ${\bf{v}}=\beta_1{\bf{v}}_1+\beta_2{\bf{v}}_2$, where $\beta_1$ and $\beta_2$ are coefficients, ${\bf{v}}_1$ and ${\bf{v}}_2$ are CA vectors\footnote{In fact, this result can be extended to multi-stream transmission, i.e., an equivalent fully-digital $N$-stream precoding can be realized provided $2N$ RF chains with CA constraints in mmWave communications \cite{sohrabi2015hybrid}.}. This result implies that by doubling the number of Tx/Rx RF chains, although CA constraint holds on each RF chain, the equivalent beamforming vector ${\bf{v}}$ does not exhibit the CA constraint any more, and thus ZF-MF without CA in the left hand side figure of Fig. \ref{fig:analog_cmp} can be realized.

\begin{figure}
\begin{minipage}[t]{0.55\linewidth}
\centering
\includegraphics[width=8 cm]{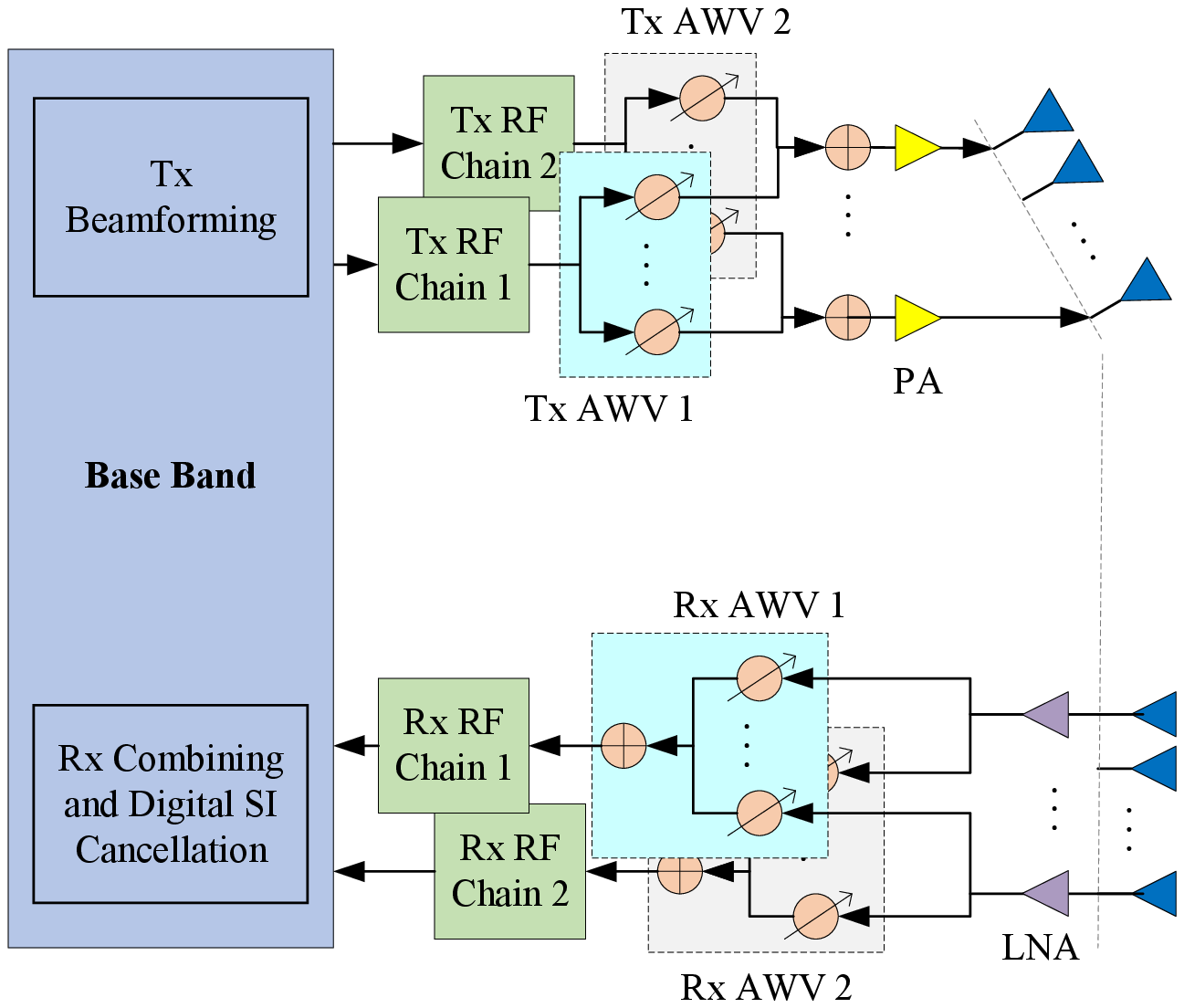}
\end{minipage}
% vfill
\begin{minipage}[t]{0.45\linewidth}
\centering
\includegraphics[width=7.4 cm]{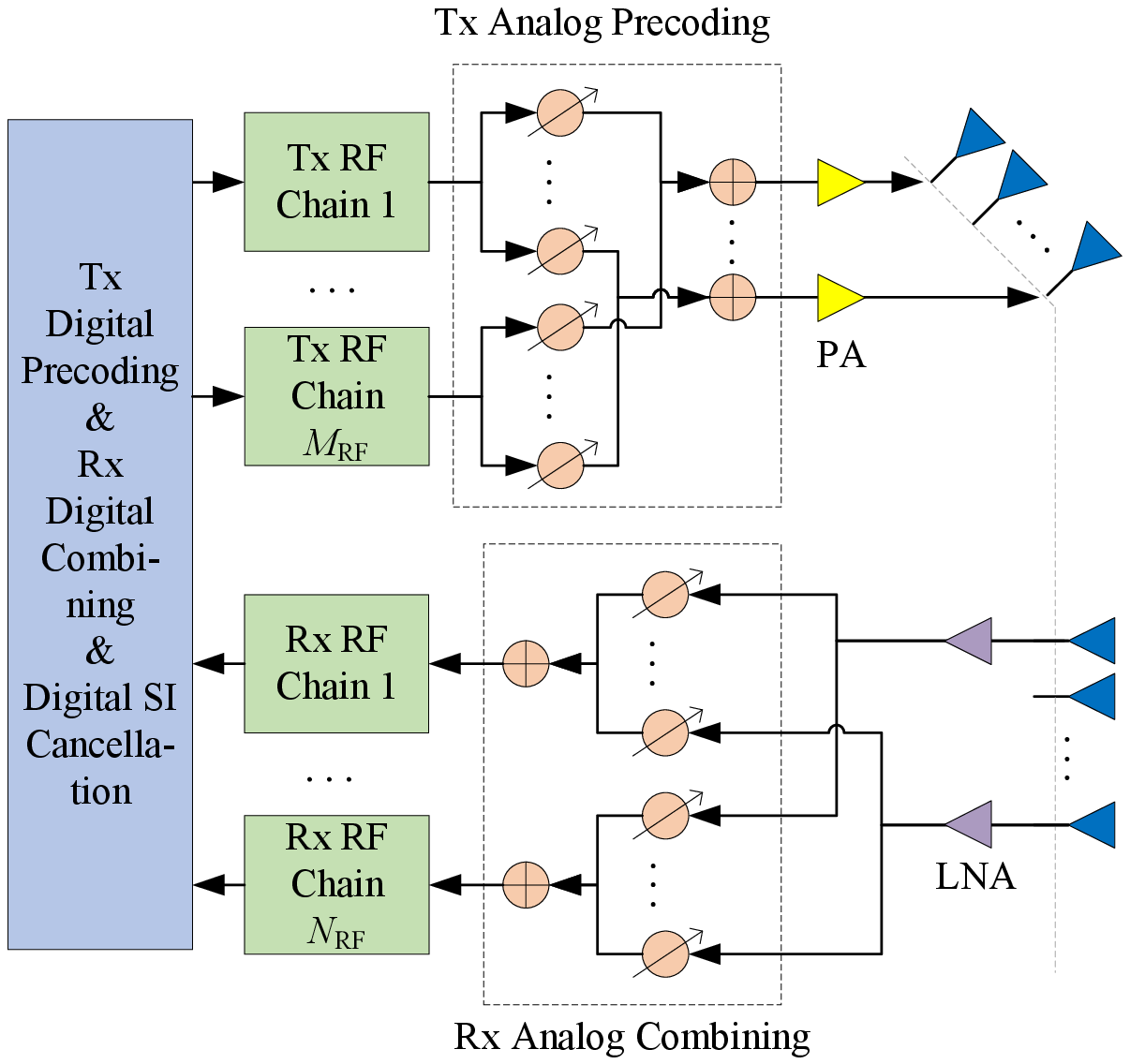}
\end{minipage}
\caption{\textbf{Left}: FD-mmWave communication with double Tx/Rx RF chains. \textbf{Right}: An FD hybrid precoding structure for the BS.}
\label{fig:hybrid}
\end{figure}

The left hand side figure of Fig. \ref{fig:hybrid} shows the structure with double Tx/Rx RF chains in an FD-mmWave node with separate Tx/Rx arrays. The Tx/Rx arrays are driven by the two independent Tx/Rx RF chains with CA AWVs. An additional Tx/Rx RF chain increases an additional degree of freedom to control the AWV, and thus the CA constraint does not appear or is circumvented in the beamforming cancellation, according to the result achieved in \cite{sohrabi2015hybrid}. On the other hand, doubling the Tx/Rx RF chains may also increase the cost and power consumptions of an FD-mmWave node. Fortunately, as shown in the figure, the two Tx/Rx RF chains can share the same up/down-convertor, power amplifiers (PAs), and low-noise amplifiers (LNAs); hence the cost and power consumptions may not increase much.

%Compared with RF cancellation, which requires an additional Tx RF chain to generate a copy of the transmitted RF signal, beamforming cancellation with double RF chains needs one more Rx RF chain. However, the performance of beamforming cancellation will be much better. Without the CA constraints, beamforming cancellation can even completely cancel the SI by using ZF filtering, including the Tx impairments (e.g., PA nonlinearity, phase noise, IQ mismatch, etc.) carried in the SI. In contrast, RF cancellation can only mitigate a part of SI, and basically

%\begin{figure}[t]
%\begin{center}
%  \includegraphics[width=10 cm]{hybrid.eps}
%  \caption{An FD hybrid precoding structure with multiple Tx/Rx RF chains.}
%  \label{fig:hybrid}
%\end{center}
%\end{figure}

%\subsection{Beamforming Training}

%\subsection{Beamforming Training}
% conventional beamforming training is not viable due, both the metric and the hierarchical search.

\section{Multi-User Scenario}
%In addition to SI, interference between different streams.
%conventional scheme is not applicable due to the CA constraint. Different approaches, compressed sensing, angle space searching.

\subsection{Benefit of FD Transmission}

In the previous section we mainly discuss point-to-point FD-mmWave transmission, which can be used for wireless backhaul link where bi-directional high data rate is required. In this section, we discuss the benefit of FD-mmWave communication in a multi-user scenario, where a base station (BS) serves multiple users. Apparently, with FD transmission the multi-user capacity can be greatly improved, because the achievable rate of each user can be almost doubled. In practice, however, this benefit may be offset, since in some cases one user does not need to transmit and receive simultaneously. For instance, when we make a call, we basically do not speak much when we are listening. Fortunately, with multiple users the BS can still sufficiently exploit the FD benefit most of the time, by receiving from some users while transmitting to some other users at the same time.

\begin{figure}
\begin{minipage}[t]{0.4\linewidth}
\centering
\includegraphics[width=6.5 cm]{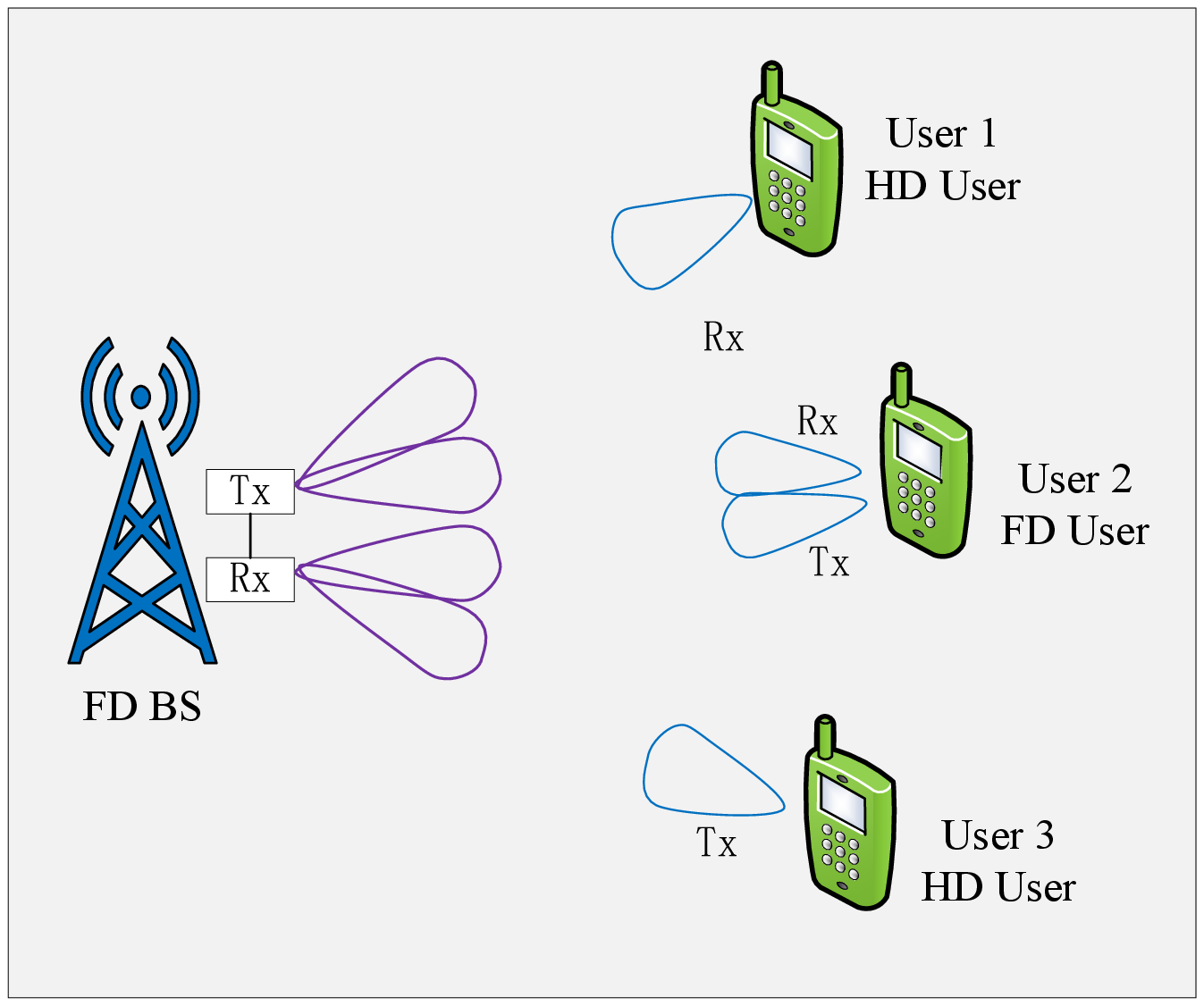}
\end{minipage}
% vfill
\begin{minipage}[t]{0.6\linewidth}
\centering
\includegraphics[width=8.5 cm]{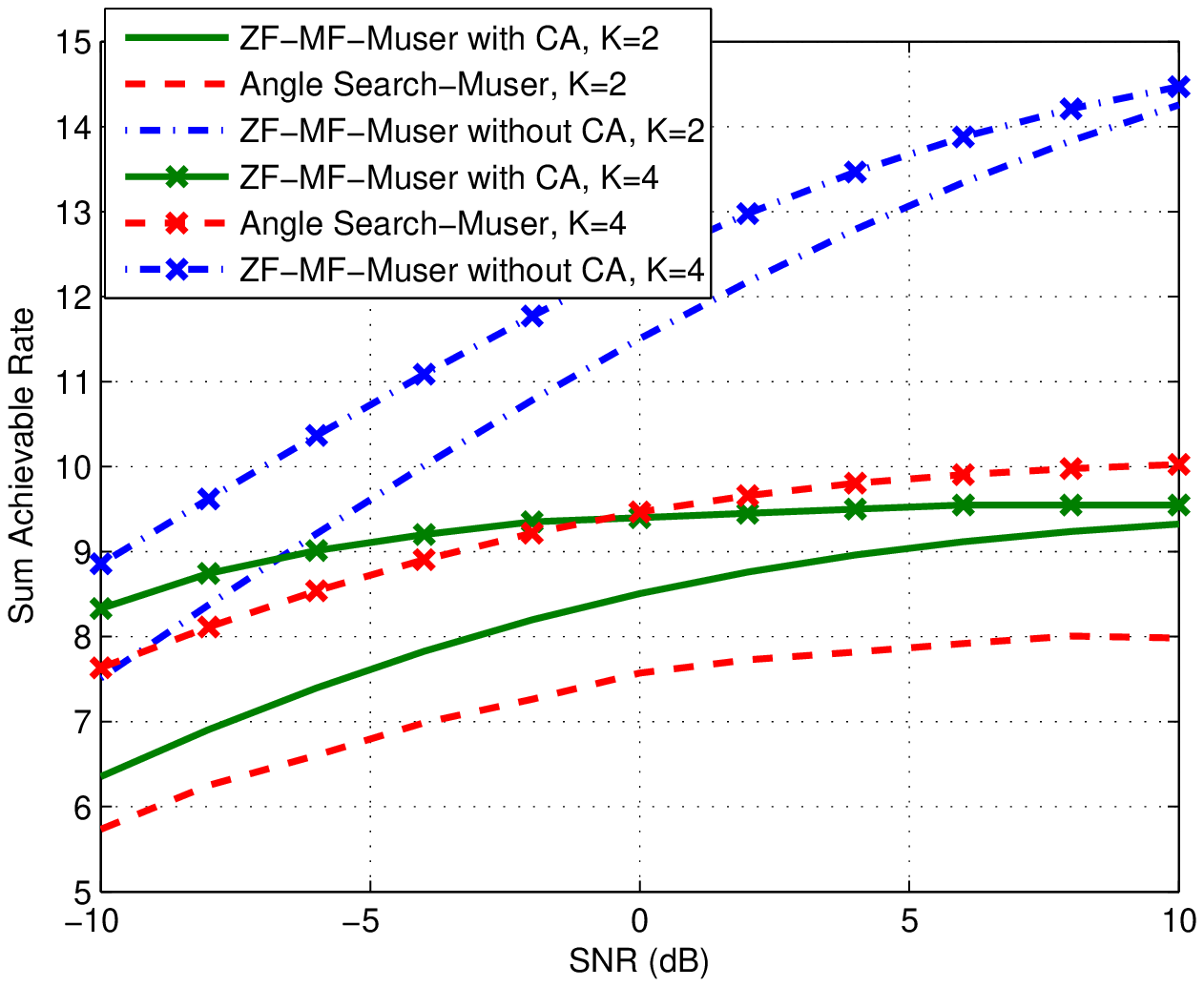}
\end{minipage}
\caption{\textbf{Left}: An FD-mmWave multi-user scenario, where an FD-mmWave BS serves multiple users, including FD users and half-duplex (HD) users. \textbf{Right}: {Achievable sum rate comparison under a multi-user FD-mmWave scenario, where one BS serves $K$ FD users. For all the users and the BS, the number of Tx and Rx antennas is 16. $\omega=\pi/8$, $d/\lambda=5$, and the SI is 20 dB w.r.t. the signal power.}}
\label{fig:multiuser}
\end{figure}

The left hand side figure of Fig. \ref{fig:multiuser} shows a typical FD-mmWave multi-user scenario, where an FD BS serves multiple users simultaneously. It is noteworthy that within these users, there are FD users, like User 2, as well as half-duplex (HD) users, like User 1 and User 3. An FD user has the antenna setting shown in Fig. \ref{fig:analog}, while an HD user can only perform conventional mmWave beamforming, either Tx or Rx, but cannot perform FD Tx/Rx beamforming. In a conventional mmWave multi-user scenario, {a mmWave} BS usually exploits a hybrid precoding structure \cite{alkhateeb2015limited} to serve multiple users with spatial division multiple access (SDMA). Thus, in an FD-mmWave multi-user scenario, an FD-mmWave BS needs to exploit an FD hybrid precoding structure shown in the right hand side figure of Fig. \ref{fig:hybrid}, where $M_{\rm{RF}}$ Tx RF chains share the same Tx array and $N_{\rm{RF}}$ Rx RF chains share the same Rx array. With this structure, the BS can transmit data to $M_{\rm{RF}}$ parallel users and meanwhile receives data from $N_{\rm{RF}}$ parallel users by using both SDMA and FD transmission.

As shown in the left hand side figure of Fig. \ref{fig:multiuser}, at the same time and the same frequency band, the BS can serve the three users simultaneously. In particular, the BS transmits data to Users 1 and 2, and distinguishes them by SDMA, meanwhile the BS receives data from Users 2 and 3, {and also distinguishes them} by SDMA. From this scenario we can find that {even if there is} no FD users in practice, the FD-mmWave BS can still exploit the FD benefit by transmitting data to some users (like User 1) while at the same time receiving data from some other users (like User 3).

\subsection{Interference Mitigation}
The FD-mmWave multi-user communication does not achieve the FD benefit for no cost. Instead, the FD-mmWave BS must face more complicated interference management. It is known that in conventional mmWave multi-user systems hybrid precoding must be exploited to mitigate multi-user interference (MUI) and meanwhile achieve array gains for multiple users \cite{alkhateeb2015limited}. In an FD-mmWave multi-user system, the BS must deal with both MUI and SI and meanwhile achieve array gains, which is called FD hybrid precoding in this paper.

Firstly, let us see the SI at the FD-mmWave BS. Unlike the point-to-point case discussed in the previous section and shown in Fig. \ref{fig:analog}, where the SI only contains one single Tx data stream, under the multi-user scenario, the SI may include multiple independent Tx data streams for multiple downlink users, which means that the SI becomes multi-dimensional. Appropriate Tx/Rx AWVs must be designed to handle the multi-dimensional SI in addition to the MUI.

On the other hand, subject to the particular antenna structure in the FD-mmWave BS, the precoding matrix (also the combining matrix) has to be decomposed into the product of a digital matrix and an analog matrix, which correspond to digital precoding and analog precoding, respectively \cite{alkhateeb2015limited}. The analog matrix has the CA constraints, which challenges the interference mitigation. In fact, in a conventional mmWave multi-user scenario, where there is only MUI but no SI, the CA constraints already make the mitigation of MUI difficult. For instance, in \cite{alkhateeb2015limited} an approach was proposed to first perform BS-user analog beamforming one by one without considering MUI in analog precoding, and then try to mitigate MUI in digital precoding where there is no CA constraint. This approach can achieve good performance, because the BS-user analog beamforming basically induces little MUI due to the spatial sparsity of the mmWave channel. For this reason, this approach may be also used for the FD hybrid precoding here. In particular, we may first perform BS-user analog beamforming by exploiting Method 1 or Method 2 in the previous section, and then design digital precoding and combining matrices to mitigate MUI and the residual SI. In fact, considering that the SI channel also has the property of spatial sparsity, as introduced in Section III, the spatially sparse precoding \cite{alkhateeb2014mimo}, which formulates the hybrid precoding problem as a sparse reconstruction problem and solves it with the principle of basis pursuit, can be also exploited to solve the FD hybrid precoding problem.

In addition, in the case that the system complexity is affordable, the method to double the number of Tx/Rx RF chains to remove the CA constraints in the previous section can be also applied in the FD-mmWave BS. In particular, the FD-mmWave BS may be equipped with $2N_{\rm{RF}}$ Rx RF chains and $2M_{\rm{RF}}$ Tx RF chains to serve $N_{\rm{RF}}$ uplink users and $M_{\rm{RF}}$ downlink users simultaneously. In such a case, the CA constraints will not appear in the beamforming cancellation according to the results in \cite{sohrabi2015hybrid}, and we can perform fully digital precoding/combining to mitigate SI and MUI, which is much more flexible than the hybrid precoding/combining. In the case when the system complexity is limited that the BS can only be equipped with $M_{\rm{RF}}$ Tx RF chains (and $N_{\rm{RF}}$ Rx RF chains), there is a tradeoff between using these RF chains to serve $M_{\rm{RF}}$ downlink users with CA constraints and serving $M_{\rm{RF}}/2$ downlink users but with the CA constraints circumvented. Channel conditions would determine which one is better.

{The right hand side figure of Fig. \ref{fig:multiuser} shows a preliminary comparison of achievable sum rate under a multi-user FD-mmWave scenario, where ZF-MF-Muser refers to the scheme that uses ZF-MF (cf. Section IV-B) between the BS and each user for SI cancellation, while Angle Search-Muser refers to the scheme that uses Angle Search (cf. Section IV-B) between the BS and each user for SI cancellation. The MUI is simply mitigated by using ZF filtering in the digital beamforming after SI cancellation. From this figure we can see that due to the CA constraint, the SI and MUI may not be completely cancelled, as there are performance floors for both ZF-MF-Muser and Angle Search-Muser at the high SNR regime. Hence, more efficient methods are in demand to deal with the SI and MUI under the CA constraint. When the CA constraint is not considered, the performance can be greatly improved, which indicates that the method of doubling the number of Tx/Rx RF chains is also effective for the multi-user scenario.}

\section{Conclusions}

In this paper, we have explored the potential of FD-mmWave communication. We have first designed antenna configurations for FD-mmWave transmission, and compared two typical settings, i.e., with separate Tx/Rx arrays or with the same array. It is shown that, different from the cases in the conventional micro-wave band FD communications, {the configuration with separate Tx/Rx arrays provides flexibility to mitigate SI with signal processing methods, while it may increase some cost and area versus that with the same array.} We then have modeled the mmWave SI channel, where a near-field propagation model was exploited to model the LOS component, and we find that the mmWave LOS-SI path also shows spatial sparsity, which is similar to the mmWave communication channel and can be used for SI mitigation.

Afterwards, as RF cancellation may not achieve satisfactory performance in mmWave communication, we have explored a new approach to mitigate SI by spatial signal processing, i.e., beamforming cancellation. With the structure constraints of mmWave beamforming taken into account, we have proposed several candidate solutions. Results show that the CA constraints lead to a significant performance loss, and thus more efficient algorithms are in demand. Moreover, the CA constraint can be circumvented in the beamforming cancellation by doubling the number of Tx/Rx RF chains, and in such a case beamforming cancellation can achieve much better performance. Lastly, we have considered an FD-mmWave multi-user scenario, and showed that even if there are no FD users in an FD-mmWave {cellular system}, the FD benefit can still be exploited in the FD BS. However, interference cancellation is more stringent in this scenario, because there are both SI and MUI. We have also proposed two candidate solutions for this scenario, which either to disregard the multi-user interference or the CA constraints first to simplify the hybrid precoding problem. In addition, the approach of doubling the number of RF chains may be also viable here provided that the complexity is affordable.

%However, if 2 Tx/Rx RF chains are available at an FD node, there is a trade off to decide whether to use the 2 RF chains for 1-stream transmission with equivalent fully-digital beamforming, or use them for 2-stream transmission with the CA constraints. After all, the former can achieve a higher SINR, but cannot achieve multiplexing gain; while the later is just opposite.

%\bibliographystyle{IEEEtran} % use IEEEtran.bst style
%\bibliography{IEEEabrv,Xiao60GHz,duplex}


\end{document}